Response to Comment on "Repulsive contact interactions make jammed particulate systems inherently nonharmonic"

Carl F. Schreck[1,2], Thibault Bertrand[1], Corey S. O'Hern[1,2,3], and Mark D. Shattuck[4,1]

[1] Department of Mechanical Engineering & Materials Science, Yale University, New Haven, CT 06520
[2] Department of Physics, Yale University, New Haven, CT 06520
[3] Department of Applied Physics, Yale University, New Haven, CT 06520
[4] Benjamin Levich Institute and Physics Department, The City College of the City of University of New York, New York, NY10031

In our Letter "Repulsive contact interactions make jammed particulate systems inherently nonharmonic" *Physical Review Letters* 107 (2011) 078301 [1], we described simulations that directly measured the vibrational response of packings of frictionless disks, instead of relying on the harmonic approximation to obtain the vibrational response as in most prior studies. We perturbed the system over a range of amplitudes and frequencies, and measured the resulting fluctuating particle positions and velocities at fixed total energy. This allowed us to compare the measured vibrational response of the packing to that inferred from linear response in the harmonic approximation. In our Letter, we employed a sensitive metric, the spectral content of the particle displacements $A_k^k$, which can distinguish parameter regimes where the vibrations are described by linear response and those that are not. We found that the critical input energy $E^*$ below which the measured vibrations of the packings are described by linear response is vanishingly small and scales as $E^* \sim \frac{(\Delta\phi)^2}{N^\beta}$, where $\beta \approx 1.7$ and $\Delta\phi$ is the amount of compression above jamming onset at packing fraction $\phi_J$. Thus, in two limits, $\Delta\phi \to 0$, for finite N and $N \to \infty$ for $\Delta\phi > 0$, linear response does not describe the measured vibrational response of these systems. This striking result stems from the fact that particulate systems with purely repulsive contact interactions continually break and reform contacts during vibrations. Linear response, which assumes a fixed contact network, cannot capture the nonlinearities that arise from contact breaking and formation.

An open discussion about the vibrational response of particulate media that undergo jamming transitions is important since the community has focused mainly on linear response for the past 15 years. In particular, most of the studies of systems that interact via finite-range, purely repulsive potentials have employed linear response to describe vibrations without determining over what range the linear approximation is valid. In their comment [2], Goodrich, Liu, and Nagel (GLN) presume that the *linear response* of jammed systems controls the behavior over a wide range of temperature and packing fraction. We find that linear response is useful at T=0, but advocate exploring new methods to understand the vibrational response of systems for T>0, where linear response fails. In contrast to conventional molecular solids that interact via Lennard-Jones or other potentials with minima for each pair separation, granular and other particulate solids interact via finite-range, purely repulsive (one-sided) potentials without minima for each pair interaction. Because of this, we argue that the vibrational response for systems that interact via finite-ranged, purely repulsive pair potentials is different from that for amorphous solids that interact via Lennard-Jones potentials (or others that possess minima for each pair interaction). In systems with one-sided interaction potentials, the breaking and forming of interparticle contacts gives rise to strong anharmonicities that are not found in conventional amorphous or crystalline solids.

To understand the breakdown of harmonic response, we note that the time-dependent particle positions $\vec{R}(t) = \{x_1, y_1, \ldots, x_N, y_N\}$ in the harmonic regime are described by

$$R_n(t) = R_n^0 + \sum_{k=1}^{2N} \tilde{R}_k \, e_{kn} \cos(\omega_k t + \phi_k), \qquad Eq.(1)$$

where $\tilde{R}_k$ are the time-independent amplitudes of the normal modes $\hat{e}_k$ with eigenfrequency $\omega_k$ from the dynamical matrix and $R_n^0$ denotes the minimum of the total potential energy. To quantify the *time-dependent* deviations of the particle positions from Eq. 1, we measure the period-averaged kinetic energy (decomposed onto normal mode p) as a function of time t following a perturbation along normal mode l,

$$K_{lp}(t = it_p) = \frac{1}{t_p} \int_{(i-1)t_p}^{t_p} \sum_{n=1}^{2N} \frac{1}{2} m (v_n e_{np})^2 \, dt, \qquad Eq.\, 2$$

where i is an integer, m is the particle mass, $v_n$ is the nth component of the velocity vector $\vec{v} = \{v_{x1}, v_{y1}, \ldots, v_{xN}, v_{yN}\}$, and $t_p = 2\pi/\omega_p$. When $2K_{ll}/K_{pert} = 1$, where $K_{pert}$ is the amplitude of the perturbation, the system is harmonic. $2K_{ll}/K_{pert} < 1$ signals deviations from harmonic behavior. $K_{ll}(\infty)$ is related to the quantity $A_k^k$ measured in our Letter [1].

Other researchers have used less sensitive methods to distinguish harmonic from nonharmonic behavior. For example, in Ref. [3] the authors argue that the linear response regime should be defined for temperatures $T<T^*$, where below $T^*$ the density of vibrational modes $D(\omega)$ becomes independent of temperature. In Ref. [3], $D(\omega)$ is the Fourier transform of the normalized velocity autocorrelation function; however, because the frequency bins are much larger than the typical mode spacing this is not a sensitive measure of harmonicity. In addition, this measure cannot detect changes in vibrational mode directions. In their comment, GLN also claim that they have investigated anharmonicities in jammed systems in Ref. [6]. However, these studies are very different from ours. They focused on zero-temperature systems and measured Grüneisen parameters defined by the variation of the eigenfrequencies of the dynamical matrix with packing fraction, not temperature. In contrast in Ref. [1], we studied packings at finite temperature T and directly compared the measured vibrational response at T>0 to that inferred from linear response of static packings at T=0.

One of the key arguments raised by GLN in their comment is their speculation that our results should depend strongly on the exponent α for the finite-ranged, purely repulsive, one-sided interaction potential

$$\frac{V(r_{ij})}{\epsilon} = \frac{1}{\alpha}\left(1 - \frac{r_{ij}}{\sigma_{ij}}\right)^\alpha \theta\left(1 - \frac{r_{ij}}{\sigma_{ij}}\right), \qquad Eq.(3)$$

where $r_{ij}$ is the separation between particle centers, $\sigma_{ij} = (\sigma_i + \sigma_j)/2$ is the average diameter, and $\theta(x)$ is the Heaviside step function. In the harmonic approximation, the particle

positions are given by Eq. (1) with eigenfrequencies $\omega_k$ from the dynamical matrix

$$M_{\alpha\beta} = \left.\frac{\partial^2 V}{\partial R_\alpha \partial R_\beta}\right|_{R=R^0}, \qquad Eq.\,(4)$$

where $R_\alpha$ indicate the particle coordinates, V is the total potential energy of the system, and *M* is evaluated at the point in configuration space where the net force on each particle is zero for a static packing at a given $\Delta\phi$. However, during vibrations, the network of interparticle contacts changes, *i.e.* contacts frequently break and near contacts form. We have shown previously that for zero-temperature packings of particles that interact via purely repulsive contact potentials the average smallest overlap (with $1-r_{ij}/\sigma_{ij} > 0$) scales as $\Delta\phi/N$, and the average smallest underlap (with $r_{ij}/\sigma_{ij} - 1 > 0$) scales as $\Delta\phi/N^2$ [2]. In Ref. [1], we found that the input energy above which the contact network changes possesses an intermediate system-size scaling. The statement by GLN, "It is only for the special case of purely repulsive Hookian springs---the case treated by SBOS---that any problem could possibly arise" is incorrect because we expect that the energy required to change the interparticle contact network will have similar system-size scaling for nearly isostatic systems that interact via purely repulsive contact potentials for all $\alpha$.

GLN argue that the system-size dependent nonlinearities identified in our Letter [1] arise from the discontinuities in the second derivatives of the interaction potential when $\alpha \leq 2$. They claim that the nonlinearities in systems with $\alpha > 2$ are not strongly system-size dependent, but are strongly time dependent. However, the nonlinearities that we discuss in Ref. [1] and this reply arise from the Heaviside step function, which represents the breaking and formation of interparticle contacts. This strong nonlinearity depends on N regardless of the value of $\alpha$.

In response to the comment by GLN, we performed molecular dynamics simulations at constant total energy for systems that interact via one-sided Hertzian spring potentials ($\alpha = 5/2$) at $\Delta\phi=10^{-4}$ and two system sizes N=12 and 48. In Fig. 1 (a), we show the normalized kinetic energy $K_{ll}$ as a function of time for single-sided Hertzian spring potentials. First, we find that for systems with one-sided Hertzian spring interactions $K_{ll}$ reaches a steady-state value at long times that decreases with the input energy. In Fig. 1 (b), we plot the long-time value of $K_{ll}$ averaged over the final 1000 periods. For repulsive linear spring potentials, $2K_{ll}^{LT}/K_{pert}=1$ until there is a rapid drop to $1/(2N-2)$ at a critical $K_{pert}$ (that decreases with increasing system size) when the first contact breaks. For one-sided Hertzian spring interactions, the behavior is similar. $K_{ll}^{LT}$ drops rapidly when a single contact breaks and the critical $K_{pert}$ decreases with increasing system size. In contrast to repulsive linear spring interactions, $K_{ll}^{LT}$ for single-sided Hertzian interactions shows a slow decrease before the first contact breaks. However, as shown in Fig. 1 (b), this nonlinearity further restricts the linear regime and becomes less important as N increases. In Fig. 1 (c), we measure the time $\tau_l$ required for $2K_{ll}$ to decay to $K_{pert}/e$ as a function of $K_{pert}$ for both one-sided linear and Hertzian springs. For input energies that do not cause contact breaking, the time scale is too large to measure for both $\alpha=2$ and 5/2. In the contact-breaking regime, $\tau_l$ decreases with increasing $K_{pert}$ for both types of interactions. Thus, we find similar strong nonharmonic behavior for systems that interact via purely repulsive contact potentials both with and without continuous second derivatives of the interaction potential at contact.

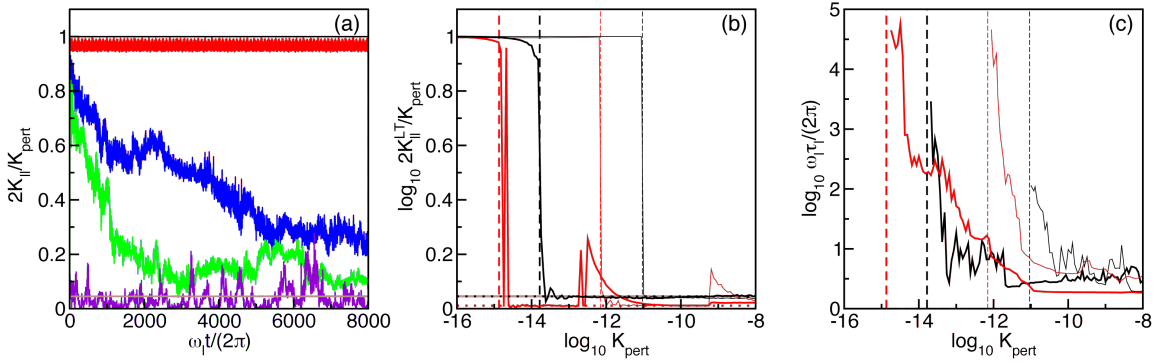

Figure 1: (a) Normalized kinetic energy $K_{ll}$ versus time $\omega_l^d \, t/(2\pi)$, in DM mode $l=2$ following a perturbation with kinetic energy $K_{pert} = 1 \times 10^{-16}$ (black), $1 \times 10^{-14}$ (red), $1.8 \times 10^{-14}$ (blue), $2.1 \times 10^{-14}$ (green), and $4.2 \times 10^{-14}$ (violet) in mode $l=2$ for a $N=12$ MS packing with Hertzian spring ($\alpha = 5/2$ in Eq. 3) interactions at $\Delta\phi = 10^{-4}$. The horizontal line indicates $2K_{ll}/K_{pert} = 1/(2N-2)$. Note that each curve begins at 1 at $t=0$. (b) Normalized kinetic energy $K_{ll}^{LT}$ in dynamical matrix mode $l$ following a perturbation with kinetic energy $K_{pert}$ in mode $l$ at long times after $\omega_l^d \, t/(2\pi) = 10^4$ cycles for systems with $N=12$ (using mode $l=2$) and 48 ($l=24$) at $\Delta\phi = 10^{-4}$. The red and black curves correspond to $N=12$ and 48, and the thin and thick curves correspond to repulsive linear ($\alpha=2$) and Hertzian spring interactions, respectively. The vertical lines correspond to the energy at which the first contact breaks for repulsive linear (dashed black) and Hertzian (dashed red) spring interactions. (c) The time scale $\tau_l$ at which the kinetic energy $2K_{ll}$ first falls below $K_{pert}/e$. The line styles and colors correspond to the same systems as in (b). Note that at input energies for which the contact network does not break, we cannot measure $\tau_l$ because $2K_{ll}$ does not decay to a value below $K_{pert}/e$.

GLN also raised a concern about the order of limits for the perturbation amplitude (displacement $\delta$ or energy E→0) and system size (N→∞). As discussed in our Letter [1], we agree that at finite N, there is a linear response regime for E< E*, where E* scales with $(\Delta\phi)^2/N^{1.7}$. However, the linear regime becomes vanishingly small and irrelevant as $\Delta\phi$→0. For the N→∞ limit, GLN claim that a single contact breaking will not strongly affect the density of vibrational modes $D(\omega)$, but we do not use $D(\omega)$ as our metric of harmonicity. In our Letter [1] and this reply, we show that the breaking of a single contact causes the rapid leakage (Fig. 1 (c)) of energy out of a pure eigenmode of the dynamical matrix of the static packing and into modes that are orthogonal to those of the dynamical matrix. The size of the characteristic energy scale that causes changes to the contact network shrinks to zero in the N→∞ limit at any $\Delta\phi$.

In their comment, GLN suggest that our statement "the density of vibrational modes cannot be described using the dynamical matrix" is tautologically incorrect. However, they confuse the phrase "normal modes of vibration" and "vibrational modes." In our Letter [1], we did not refer to the "density of normal modes," instead we referred to the "density of vibrational modes". It is important to distinguish the measured vibrational response of a system, *i.e.* the frequency content of a fluctuating system undergoing real dynamics, from approximate expressions for the vibrational response of static packings, *e.g.* normal modes. In our Letter and this reply, the phrase "density of vibrational modes" refers to the frequency content of the measured vibrational response of the system.

GLN claim that it is well established that anharmonic effects dominate at the jamming transition and cite Refs. [3,5,6,7]. This statement is extremely misleading. Refs. [3] and [7] were published after our Letter appeared. Refs. [5] and [6] only infer anharmonic effects from the eigenmodes of the dynamical matrix of static packings. Neither Ref. [5] nor [6] measure the vibrational response of static packings at finite temperature. In addition, GLN argue that it is essential to understand how linear quantities behave near jamming since the transition controls the vibrational response at larger values of compression. However, for small $\Delta\phi$, the linear regime is vanishingly small, and thus the linear theory cannot describe systems that exist near jamming at $\phi_J$. We advocate focusing on the larger, more important nonlinear response of jammed systems near, not far above $\phi_J$.

We agree with GLN that the "jamming transition" controls vibrations at large as well as small $\Delta\phi$ for input energies $E < E^* \approx (\Delta\phi)^2/N^{1.7}$. However, we have shown in preliminary studies [8] that most of the $\phi$ and T phase diagram where contacts can break and form and below the energy where cage-breaking rearrangements occur has a common vibrational response, which is significantly different from the linear vibrational response near jamming.